\documentclass[trackchanges,preprint2]{aastex701}

\usepackage{array} 
\usepackage{multirow}
\usepackage{amsmath}
\newcolumntype{M}[1]{>{\centering\arraybackslash}m{#1}}

\begin{document}

\title{The JWST Rocky Worlds DDT Program reveals GJ\,3929b to likely be a bare rock}

\author[orcid=0000-0002-6215-5425]{Qiao Xue}
\affiliation{Department of Astronomy and Astrophysics, University of Chicago, Chicago, IL 60637, USA}
\email{qiaox@uchicago.edu}

\author[orcid=0000-0002-0659-1783]{Michael Zhang}
\altaffiliation{51 Pegasi b Postdoctoral Fellow}
\affiliation{Department of Astronomy and Astrophysics, University of Chicago, Chicago, IL 60637, USA}
\email{zmzhang@uchicago.edu}

\author[orcid=0000-0002-0508-857X]{Brandon Park Coy}
\affiliation{Department of the Geophysical Sciences, University of Chicago, Chicago, IL 60637, USA}
\email{bpcoy@uchicago.edu}

\author[orcid=0000-0003-2404-2427]{Madison Brady}
\affiliation{Department of Astronomy and Astrophysics, University of Chicago, Chicago, IL 60637, USA}
\email{mtbrady@uchicago.edu}

\author[orcid=0000-0002-1592-7832]{Xuan Ji}
\affiliation{Department of the Geophysical Sciences, University of Chicago, Chicago, IL 60637, USA}
\email{xuanji@uchicago.edu}

\author[orcid=0000-0003-4733-6532]{Jacob L.\ Bean}
\affiliation{Department of Astronomy and Astrophysics, University of Chicago, Chicago, IL 60637, USA}
\email{jacobbean@uchicago.edu}

\author[orcid=0000-0002-3328-1203]{Michael Radica}\thanks{NSERC Postdoctoral Fellow}
\affiliation{Department of Astronomy and Astrophysics, University of Chicago, Chicago, IL 60637, USA}
\email{radicamc@uchicago.edu}

\author[0000-0003-4526-3747]{Andreas Seifahrt}
\affiliation{Gemini Observatory/NSF NOIRLab, 670 N. A'ohoku Place, Hilo, HI 96720, USA}
 \email{andreas.seifahrt@noirlab.edu}
 
\author[0000-0002-4410-4712]{Julian St\"urmer}
\affiliation{Landessternwarte, Zentrum f\"ur Astronomie der Universit\"at Heidelberg, K\"onigstuhl 12, D-69117 Heidelberg, Germany}
\email{stuermer@lsw.uni-heidelberg.de}

\author[orcid=0000-0002-4671-2957]{Rafael Luque} \thanks{NHFP Sagan Fellow}
\affiliation{Department of Astronomy and Astrophysics, University of Chicago, Chicago, IL 60637, USA}
\affiliation{Instituto de Astrof\'isica de Andaluc\'ia (IAA-CSIC), Glorieta de la Astronom\'ia s/n, 18008 Granada, Spain}
\email{rluque@uchicago.edu}

\author[0000-0003-4508-2436]{Ritvik Basant}
\affiliation{Department of Astronomy and Astrophysics, University of Chicago, Chicago, IL 60637, USA}
\email{rbasant@uchicago.edu}

\author[0009-0003-1142-292X]{Nina Brown}
\affiliation{Department of Astronomy and Astrophysics, University of Chicago, Chicago, IL 60637, USA}
\email{ninabrown@uchicago.edu}

\author[0009-0005-1486-8374]{Tanya Das}
\affiliation{Department of Astronomy and Astrophysics, University of Chicago, Chicago, IL 60637, USA}
\email{tanyadas@uchicago.edu}

\author[0000-0003-0534-6388]{David Kasper}
\affiliation{Department of Astronomy and Astrophysics, University of Chicago, Chicago, IL 60637, USA}
\email{kasp0127@gmail.com}

\author[orcid=0000-0002-2875-917X]{Caroline~Piaulet-Ghorayeb}\thanks{E.\ Margaret Burbridge Postdoctoral Fellow}
\affiliation{Department of Astronomy and Astrophysics, University of Chicago, Chicago, IL 60637, USA}
\email{carolinepiaulet@uchicago.edu}

\author[orcid=0000-0002-1337-9051]{Eliza M.-R.\ Kempton}
\affiliation{Department of Astronomy and Astrophysics, University of Chicago, Chicago, IL 60637, USA}
\affiliation{Department of Astronomy, University of Maryland, College Park, MD 20742, USA}
\email{ekempton@uchicago.edu}

\author[orcid=0000-0002-1426-1186]{Edwin Kite}
\affiliation{Department of the Geophysical Sciences, University of Chicago, Chicago, IL 60637, USA}
\email{kite@uchicago.edu}

\begin{abstract}
We report first results from the JWST Rocky Worlds Director's Discretionary Time program. Two secondary eclipses of the terrestrial exoplanet GJ\,3929b were recently observed using MIRI photometric imaging at 15\,$\mathrm{\mu m}$. We present a reduction of these data using the updated \texttt{SPARTA} pipeline. We also refine the planet mass, radius, and predicted time of secondary eclipse using a new sector of TESS data and new, high-precision radial velocities from the MAROON-X spectrograph. For the two JWST observations, we recover secondary eclipse depths of 177\,$^{+47}_{-45}$\,ppm and 143\,$^{+34}_{-35}$\,ppm at times consistent with a nearly circular orbit, as expected from the radial velocity data. A joint fit of the two visits yields a dayside brightness temperature $T_{p,\mathrm{dayside}} = 782\pm79$\,K for GJ\,3929b, which is consistent with the maximum brightness temperature $T_{\mathrm{max}} = 737 \pm 14$\,K for a bare, black rock (i.e., assuming zero Bond albedo and no heat redistribution). These results rule out CO$_2$-rich atmospheres thicker than 100\,mbar at $>3\sigma$, suggesting that GJ\,3929b has lost any significant secondary atmosphere. The radial velocity data also indicate two additional non-transiting planets in the system: a previously-identified planet in a 15.0\,d orbit, and a newly-identified planet candidate in a 6.1\,d orbit.



\end{abstract}



\section{Introduction}\label{sec:introduction}
A major goal of NASA's James Webb Space Telescope (JWST), as well as NASA's missions in planning, such as the Habitable Worlds Observatory (HWO), is to determine the prevalence and origins of terrestrial (i.e., smaller than 1.4R$_\mathrm{_\oplus}$) exoplanet atmospheres. To achieve this goal, it is essential to study terrestrial exoplanets orbiting M dwarfs. Since approximately 75\% of all main-sequence stars in the Milky Way are M dwarfs \citep{henry_solar_2018,reyle_10_2021}, and M dwarfs host more rocky planets than Sun-like stars, those orbiting M dwarfs are the most abundant rocky planets in the galaxy \citep{dressing_occurrence_2015,mulders_stellar-mass-dependent_2015}. Additionally, because the habitable zone around smaller, cooler stars lies much closer in, temperate planets have shorter orbital periods and a higher probability of transiting. For a given planet size, the smaller stellar radius also makes their transits deeper. All these factors make rocky planets around M dwarfs the most observationally favorable targets for studying terrestrial exoplanet atmospheres with current facilities, paving the way to understanding planetary habitability.

Despite the observationally favorable traits of their planets, M dwarf stars pose a formidable threat to planetary atmospheres. They emit excess UV and X-ray radiation during their extended pre-main-sequence phase \citep{davenport_multi-wavelength_2012}, driving thermal escape that can strip volatiles \citep{ehrenreich_massloss_2011,ehlman_sustainability_2016,gronoff_atmospheric_2019}. Even on the main sequence, frequent flares and strong stellar winds continue to erode atmospheres \citep{davenport_multi-wavelength_2012}. 

Atmospheric loss processes give rise to the theoretical ``cosmic shoreline'' -- the concept that a planet’s ability to maintain an atmosphere depends on its escape velocity (v$_\mathrm{esc}$) and its cumulative XUV irradiation (I$_\mathrm{XUV}$). In our solar system, the empirically-derived shoreline follows the relation $\mathrm{I_{XUV}}\sim \mathrm{v_{esc}^4}$, and divides bodies with atmospheres from those without \citep{zahnle_cosmic_2017}. This shoreline concept has been adopted among the exoplanet community to identify rocky exoplanets most likely to have atmospheres versus those not.

Based on updated historic XUV fluence estimates from real observations of mid- to late-M dwarfs, \citet{pass_receding_2025} propose a more pessimistic cosmic shoreline, with planets orbiting late-type stars less likely to retain atmospheres than predicted by the original model. More recent work by \citet{berta-thompson_3d_2025} used a data-driven Bayesian framework, demonstrating a stellar luminosity-dependent shoreline that leaves planets around low-luminosity M dwarfs especially vulnerable to atmospheric loss. However, \citet{ji_cosmic_2025} show that the shoreline predicted by detailed hydrodynamic escape models deviates from the simple scaling law, implying that super-Earths are more likely to retain an atmosphere than what traditional shoreline models predicted. In their work, the shoreline is not a sharp line but a broad region influenced by factors such as a planet’s initial volatile inventory and atmospheric composition.

Close-in planets are expected to be tidally locked, with a warm permanent dayside that always faces the star and a colder unilluminated nightside. 
Among the techniques used to study the atmospheres of transiting exoplanets, measuring the thermal emission from a planet’s dayside during secondary eclipse is a particularly effective way to test whether a tidally locked terrestrial planet has an atmosphere \citep{koll_eclipse_2019, mansfield_identifying_2019}. This method is less expensive than full-orbit observations and is not affected by the Transit Light Source Effect (TLSE) — stellar surface inhomogeneities that can bias atmospheric measurements \citep{rackham_transit_2018, lim_atmospheric_2023,radica_promise_2024}.

The key idea behind using the secondary eclipse technique to probe the presence of atmospheres is that thick atmospheres redistribute heat from the dayside to the nightside more efficiently than thin atmospheres or bare surfaces \citep{koll_eclipse_2019, koll_scaling_2022}. While some types of fresh regolith have the potential for high Bond albedo that would lead to lower temperatures on the dayside \citep{hu12,paragas25,hammond_2025}, long-term processes like space weathering are thought to efficiently lower the Bond albedo of these surfaces (e.g., \citealt{lyu2024,coy2025}).  Therefore, almost regardless of composition, the dayside of a rocky planet without an atmosphere is expected to be nearly as hot as the theoretical maximum. Additionally, atmospheres themselves could have higher albedo than bare rock surfaces due to reflective clouds \citep{mansfield_identifying_2019}. Ultimately, observing a cooler dayside than expected for a ``bare, black rock'' would suggest the presence of an atmosphere. 

The first observational studies with thermal emission suggest that thick atmospheres are not common on warm-to-hot rocky planets around M dwarfs. Spitzer thermal emission photometry indicated bare-rock surfaces for LHS\,3844b and GJ\,1252b \citep{kreidberg_absence_2019,crossfield_gj_2022}. Early JWST studies using MIRI/LRS (GJ\,367b: \citealt{zhang_gj_2024}, GJ\,1132b: \citealt{xue_jwst_2024}, GJ\,486b: \citealt{weiner_mansfield_no_2024},  LTT\,1445Ab: \citealt{wachiraphan_thermal_2024}), MIRI/Imaging with F1280W and F1500W (TRAPPIST-1b: \citealt{greene_thermal_2023,ducrot_combined_2024}, TRAPPIST-1c: \citealt{zieba_no_2023}, TOI-1468b: \citealt{valdes_hot_2025}, LHS\,1140c: \citealt{fortune_hot_2025}), and  NIRSpec/G395H (TOI-1685b, \citealt{luque_dark_2025}) likewise reveal dayside brightness temperatures consistent with minimal atmospheres. These planets span a wide range of equilibrium temperatures, from temperatures hot enough to melt the majority of their dayside surfaces (GJ\,367b, $T_{eq}$\,=\,1,365\,K) to temperatures similar to that of Venus (TRAPPIST-1c, $T_{eq}$\,=\,340\,K), suggesting that atmospheric loss is widespread and efficient for M-Earths. LHS\,1478b is the only exception, showing a shallower eclipse that could indicate an atmosphere \citep{august_hot_2025}, but the data are dominated by systematics at the level of the expected signal, requiring confirmation with future observations \citep{august_jwstprop_2025}. 

To expand the empirical base for understanding M-Earth atmospheres, 500 hours of JWST Director’s Discretionary Time (DDT) have been allocated to the Rocky Worlds DDT program \citep{redfield_report_2024}. The first selected target in this program is GJ\,3929b, a warm ($T_{eq}=568\,$K) terrestrial planet on a 2.62-day orbit around a nearby M3.5V star \citep{kemmer_discovery_2022,beard_gj_2022}. GJ\,3929b has an Emission Spectroscopy Metric (ESM) of $\sim$4.1 \citep{kempton_framework_2018}, making it a high-priority candidate for thermal emission studies. 

In this Letter, we report results from the first two visits of GJ\,3929b from the Rocky Worlds DDT. We concentrate on two unique contributions that our team can make to this public survey. The first is an analysis of new MAROON-X radial velocities for the system that improves the precision on the planet mass and the expected time of secondary eclipse (\S\ref{sec:mx}). The second is data reduction using our newly updated \texttt{SPARTA} pipeline (\S\ref{sec:jwst}). In \S\ref{sec:model}, we interpret the results in terms of potential atmospheric and surface properties. We discuss the results in the context of the cosmic shoreline in \S\ref{sec:discussion}, and we conclude in \S\ref{sec:conclusion}.

\section{Improved parameters from new TESS and MAROON-X data}\label{sec:mx}

\begin{table*}[!t]
\begin{tabular}{cl|cc}
\hline
\multicolumn{2}{l|}{\textbf{Parameter}}                                  & \textbf{Prior} & \textbf{Posterior} \\ \hline
\multirow{18}{*}{}       & $P_{b}$ (d)                                   & $\mathcal{N}$(2.61625076, 0.0000501)      & $2.6162644^{+2.3e-06}_{-2.6e-06}$   \\
                         & $K_{b}$ (m\,s$^{-1}$)                         & $\mathcal{U}$(0, 20)                      & $1.16^{+0.08}_{-0.09}$           \\
                         & $T_{0,b}$ (BJD)$^*$                               & $\mathcal{N}$(2460452.88931472, 0.1)  & $2460452.8997^{+0.0011}_{-0.0012}$  \\
                         & $e_{b}$                                       & $\mathcal{\beta}$(1.78, 9.43)             & $0.043^{+0.030}_{-0.021}$           \\
                         & $\omega_{b}$ (degrees)                        &  $\mathcal{U}$(-180, 180)                 & $-131^{+71}_{-86}$         \\
                         & $r_{1,b}$                                     &  $\mathcal{U}$(0, 1)                      & $0.48^{+0.12}_{-0.09}$           \\
                         & $r_{2,b}$                                     &  $\mathcal{U}$(0, 1)                      & $0.0318^{+0.0007}_{-0.0006}$        \\[3mm]
                         & $P_{c}$ (d)                                   & $\mathcal{N}$(15.0, 1.0)                  & $14.994^{+0.008}_{-0.008}$       \\
                         & $K_{c}$ (m\,s$^{-1}$)                         &  $\mathcal{U}$(0, 20)                     & $2.75^{+0.17}_{-0.16}$              \\
                         & $T_{0,c}$ (BJD)                                &  $\mathcal{U}$(2459374.0, 2459390.0)     & $2459386.41^{+0.21}_{-0.23}$         \\
                         & $e_{c}$                                       & $\beta$(1.78, 9.43)             & $0.16^{+0.05}_{-0.04}$           \\
                         & $\omega_{c}$ (degrees)                        &  $\mathcal{U}$(-180, 180)                  & $85^{+13}_{-20}$           \\[3mm]
                         & $P_{d}$ (d)                                   & $\mathcal{N}$(6.1, 1.0)                   & $6.116\,\pm\,0.004$        \\
                         & $K_{d}$ (m\,s$^{-1}$)                         &  $\mathcal{U}$(0, 20)                      & $1.06\,\pm\,0.11$              \\
                         & $T_{0,d}$ (BJD)                              &  $\mathcal{U}$(2459374.0, 2459381.0)        & $2459379.63\,\pm\,0.12$        \\
                         & $e_{d}$                                       & $\mathcal{\beta}$(1.78, 9.43)             & $0.13^{+0.05}_{-0.06}$           \\
                         & $\omega_d$ (degrees)                          &  $\mathcal{U}$(-180, 180)                  & $68^{+45}_{-39}$           \\[3mm]
                         & $\rho_\star$ (kg\,m$^{-3}$)                   & $\mathcal{N}$(12496, 1390)                & $13693^{+1061}_{-792}$             \\ \hline
\multirow{6}{*}{\textbf{Derived}} & $M_{b}$ ($M_\oplus$)                          & \multicolumn{2}{c}{$1.13\,\pm\,0.09$}                                 \\
                         & $R_b/R_\star$                                 & \multicolumn{2}{c}{$0.0318^{+0.0007}_{-0.0006}$}                      \\
                         & $a_b/R_\star$                                 & \multicolumn{2}{c}{$17.05^{+0.43}_{-0.34}$} 
                         \\
                         & $i_{b}$ (degrees)                             & \multicolumn{2}{c}{$89.3^{+0.4}_{-0.6}$}                           \\
                         & $R_{b}$ ($R_\oplus$)                          & \multicolumn{2}{c}{$1.09\,\pm\,0.04$}                                 \\[3mm]
                         & $m$\,sin\,$i_{c}$ ($M_\oplus$) & \multicolumn{2}{c}{$4.76^{+0.34}_{-0.32}$}                                 \\
                         & $m$\,sin\,$i_{d}$ ($M_\oplus$) & \multicolumn{2}{c}{$1.38^{+0.15}_{-0.14}$}                            \\ \hline
\end{tabular}
    \caption{Priors and posteriors of the joint fit to the GJ\,3929 RVs and transit data. Parameters are defined according to the \texttt{juliet} package \citep{juliet}. $\mathcal{U}$ denotes a uniform prior between the specified lower and upper bounds; $\mathcal{N}$ denotes a normal prior, where the first value is the mean and the second is the standard deviation. We also report $T_{0,b}$=$2459241.56931\pm0.00028$\,BJD chosen to minimize its covariance with the orbital period.}
    \label{tab:GJ3929_joint_fit}
\end{table*}

A precise prediction of the secondary eclipse time of GJ\,3929b is essential to capture the event in the JWST data, and robust mass and radius measurements are crucial to informing the interior composition and atmospheric properties. CARMENES \citep{quirrenbach_carmenes_2014} radial velocities (RVs) on the target in \citet{kemmer_discovery_2022} yielded a planetary mass of 1.21$^{+0.40}_{-0.42}$\,M$_\oplus$, a 33\% uncertainty that is too large to differentiate among plausible bulk compositions. A subsequent RV analysis by \citet{beard_gj_2022}, using the CARMENES data from \citet{kemmer_discovery_2022} in addition to RVs from NEID \citep{schwab_neid_2016} and HPF \citep{mahadevan_hpf_2012}, reported a substantially higher mass of 1.75$^{+0.44}_{-0.45}$\,M$_\oplus$, but again with large uncertainties. Furthermore, because neither analysis detected the planet to high confidence, the literature constraints on the eccentricity and longitude of periastron are insufficient to predict the secondary eclipse time when accounting for the possibility of an elliptical orbit.

GJ\,3929b is a member of the 30\,pc HUMDRUM sample of transiting planets orbiting M dwarfs \citep{brady_humdrum_2024} that are being targeted using the MAROON-X spectrograph \citep{seifahrt_development_2016, seifahrt_maroon-x_2018, seifahrt_-sky_2020, seifahrt_maroon-x_2022}. As part of this survey, we observed GJ\,3929b 77 times between February 2021 and July 2022 (covering two observing seasons). The typical exposure time was 900\,s and the derived radial velocities have an average precision of 86 and 48\,cm\,s$^{-1}$ in the blue- and red-arm data, respectively. We provided these data to the Rocky Worlds Core implementation team in December 2024 to aid the scheduling of the JWST observations.

For this paper, we analyzed the MAROON-X radial velocities of GJ\,3929 following the procedure outlined in \citet{brady_humdrum_2024}. We used the calibrated offsets for the MAROON-X data from \cite{basant_calibrations_2025}, and performed a joint fit of the transit and radial velocity data using \texttt{juliet} \citep{juliet}. We used transit data from TESS \citep{ricker_transiting_2014} sectors 24, 25, and 78 (the latter is new since the initial discovery) that were extracted using \texttt{lightkurve} \citep{lightkurve}. After inspection of the publicly available ground-based light curves, we also elected to include the data from ARCTIC observations on February 26 and April 30, 2021 (Astrophysical Research Consortium (ARC) Telescope Imaging Camera; \citealt{huehnerhoff_astrophysical_2016}), and a MuSCAT3 observation from April 15, 2021 (Multi-color Simultaneous Camera for studying Atmospheres of Transiting exoplanets 3 camera; \citealt{MuSCAT3_narita_2020}) as reported in \citet{beard_gj_2022}. We included radial velocity data from the literature obtained with the CARMENES, NEID, and HPF instruments \citep{kemmer_discovery_2022, beard_gj_2022}. 

To select our final model for the joint fit with the transits, we first performed an analysis of the RV data alone using the $\ell_1$ periodogram described in \citep{hara_l1_2017}.  The $\ell_1$ periodogram takes in user-defined noise parameters and produces a sparse representation of the signals present in RV data.  After examining a grid of noise parameters (described in more detail in M.\ Brady et al.\ in prep), we found strong evidence (log$_{10}$\,FAP\,$<$\,$-$6) of the transiting planet and an additional signal at around 15.0\,d.  This 15.0\,d signal was previously identified by \citet{kemmer_discovery_2022} and \citet{beard_gj_2022}, though it was dismissed as an alias of a planet with a true period of 14.3d in \citet{kemmer_discovery_2022}.  Given the increased significance of the 15.0\,d signal after the addition of the MAROON-X data, we believe this to be the true period of GJ\,3929\,c.   We also found log$_{10}$\,FAP\,$<$\,$-$3 for a third, 6.1\,d signal, but this signal was not significant in all $\ell_1$-periodogram noise realizations.  We included the 15.0\,d and 6.1\,d signals as Keplerians in our \texttt{juliet} fits, as well as a Double Simple Harmonic Oscillator GP \citep{kossakowski_gp_2022} signal with a period of 122\,$\pm$\,13\,d meant to reflect the stellar rotation period measured in \cite{kemmer_discovery_2022}.  We studied two classes of models- one in which the planets’ eccentricities were fixed at zero, and one in which the planets’ eccentricities were allowed to vary, with a prior based on the Beta distribution of eccentricities for hot-super Earths from \citet{stevenson_eccentricity_2025}.

\begin{figure*}[!t]
    \centering
    \includegraphics[width=0.98\linewidth]{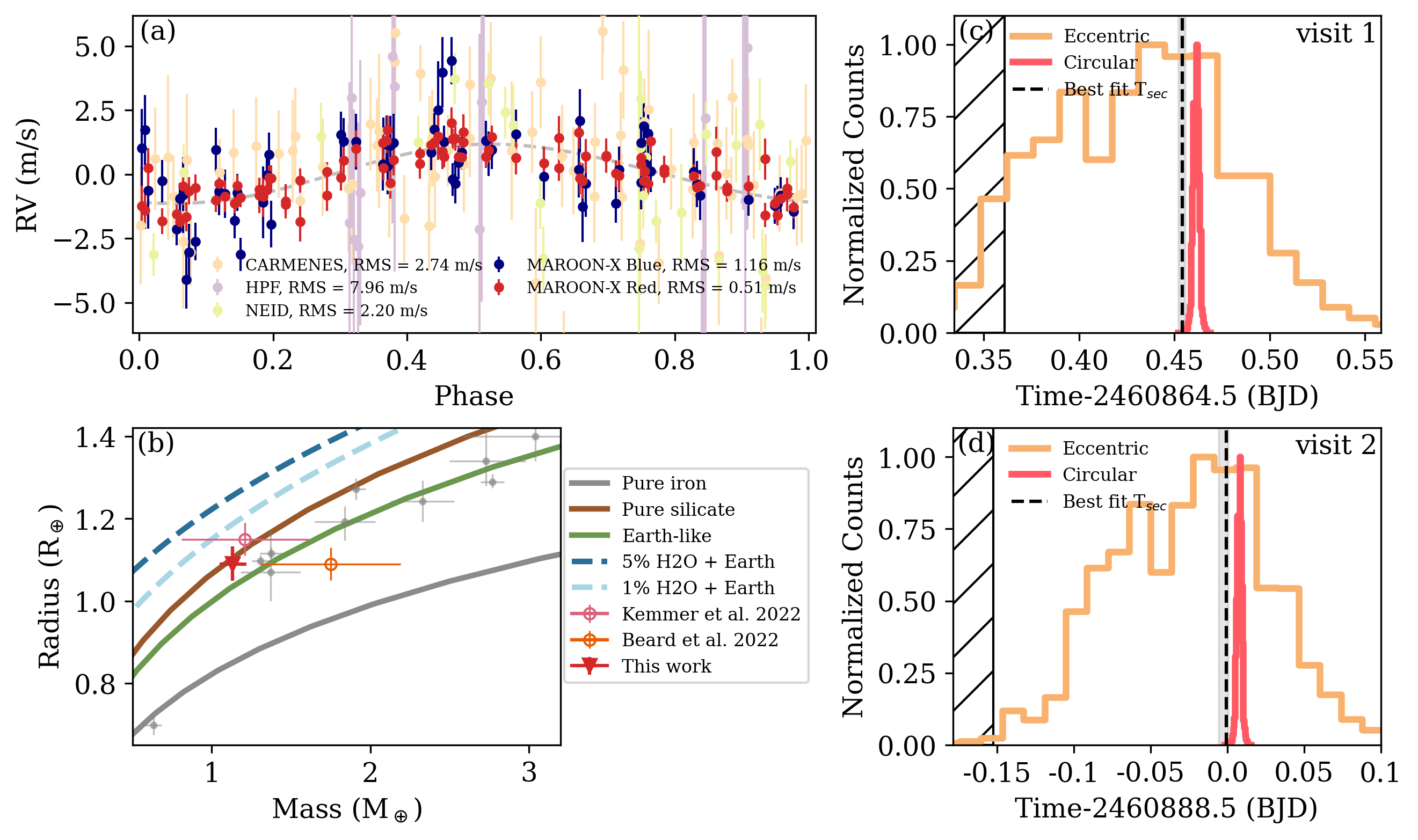}
    \caption{\textbf{(a)}: Phase-folded RV measurements from MAROON-X, CARMENES, HPF, and NEID used in our joint fit, overplotted with the best-fit eccentric-orbit model with rotation GP. \textbf{(b)}: GJ\,3929b in the mass–radius diagram with $R_p$ and $M_p$ from the joint fit (see \S \ref{sec:mx}). With the new MAROON-X data, we are able to improve the mass precision by a factor of 3.5. We compare the planet's mass and radius to various interior composition models from \citet{zeng_growth_2019}, as well as steam atmosphere models with 1\% and 5\% water from \citet{turbet_revised_2020}. GJ\,3929b is most consistent with a purely rocky or Earth-like composition (32.5\% Fe+67.5\% MgSiO$_3$). The grey points are all the other rocky planets with emission measurements from JWST. \textbf{(c) and (d)}: Median and $\pm1\sigma$ uncertainty of the measured secondary eclipse time for each individual visit compared to the predicted T$_\mathrm{sec}$ from our joint RV and transit fit. The prediction is T$_\mathrm{sec}$ = T$_0 + \frac{P}{2}(1+\frac{4}{\pi} e \mathrm{cos}\omega)+\mathrm{N}P + \frac{2a}{c}$, where T$_0$, $P$, $e$, and $\omega$ are drawn from the joint fit chains, N is the number of epochs between the reference epoch T$_0$ and the JWST observation, and $\frac{2a}{c}$ accounts for the light travel time. The circular orbit predictions are obtained by fixing $e$ to 0. Covariance between T$_0$ and $P$ is fully propagated by sampling directly from the joint‐fit MCMC chains. The x-axis shows the full observation window, with the shaded region indicating the part excluded from the fitting.}
    \label{fig:RV_mass}
\end{figure*}

Overall, we found strong ($\Delta$ln\,$Z\,>\,5$) evidence for each of the three Keplerians, as well as the rotation GP, in our data.  However, we only found weak-to-moderate evidence ($\Delta$ln\,$Z\,\approx\,3$) for non-circular orbits in this four-signal model.  However, we used the model with varying eccentricities in our final joint fits with the transit data because we are interested in constraining the secondary eclipse time of GJ\,3929b in the face of a possibly elliptical orbit. The results of our joint RV/transit fits to this model with \texttt{juliet} are shown in Table~\ref{tab:GJ3929_joint_fit}, and a phase-folded radial velocity fit for GJ\,3929b and mass-radius diagram are shown in Figure~\ref{fig:RV_mass}. We adopted the stellar parameters from \citet{kemmer_discovery_2022} after checking them with an analysis of our MAROON-X spectra \citep[following the approach of][]{brady_humdrum_2024}.

We find a mass of 1.13\,$\pm$\,0.09\,M$_{\oplus}$ for GJ\,3929b. The addition of the MAROON-X data pushes the precision on the measurement of the velocity semi-amplitude to more than 12$\sigma$, which yields a 8\% measurement precision on the planet mass when including the uncertainty on the stellar mass. We also revise the planet radius to 1.09\,$\pm$\,0.04\,R$_{\oplus}$. This places the planet slightly above the Earth-like composition line in the mass-radius diagram. Using the code \texttt{smint} \citep{piaulet_wasp107_2021, piaulet_gj9827_2024}, which utilizes the rocky planet composition models from \cite{zeng_rocky_2016}, we estimated the planet's core mass fraction (CMF).  We found that our revised mass and radius lead to a planetary CMF of $0.21^{+0.12}_{-0.16}$, which is lower than but still $1\sigma$ consistent with that of the Earth \citep[0.33, see][]{szurgot_cmf_2015}. 

From the RV and transit joint analysis, we determine an eccentricity of $0.043^{+0.030}_{-0.021}$ and $e$\,cos$\,\omega= -0.014^{+0.028}_{-0.034}$ for GJ\,3929b. The expected secondary-eclipse mid-time is shifted earlier than phase 0.5 by 0.5\,h, with a 2$\sigma$ window of 1.9\,hr. However, a circular orbit is still consistent with the RV data at the 2$\sigma$ level.

\section{JWST characterization}\label{sec:jwst}
\subsection{Observations}
JWST observed GJ\,3929b around predicted times of secondary eclipse on 2025 July 8 and 2025 July 31 as part of program DD 9235 \citep{2025jwst.prop.9235E}. Each visit was comprised of MIRI time-series imaging using the 1500W filter and the SUB128 array. The detector was read using the FAST1R mode with 47 groups per integration. The first visit had 3,387 integrations, covering phases 0.45 -- 0.54 (5.4\,hr of data). The second visit had 3,863 integrations, covering phases 0.43 -- 0.53 (6.1\,hr of data). For reference, the eclipse duration is 1.2\,h.

Each secondary eclipse visit was preceded by a brief (4 groups/integration) pre-slew observation of the field the telescope was pointed at before the science target using the MIRI prism (denoted P750L). The purpose of this observation was to assess whether the filter that is used prior to the program affects the time-dependent detector settling slope seen at the beginning of the science exposures \citep[][and private communication from STScI help desk]{fortune_hot_2025}. We find very different ramps at the beginning of the two visits (a decreasing ramp for visit 1 and an increasing ramp for visit 2), which suggests that pre-flashing the detector with a short prism exposure does not homogenize the ramps.


\subsection{Data reduction}

\begin{figure}
    \centering
    \includegraphics[width=1.\linewidth]{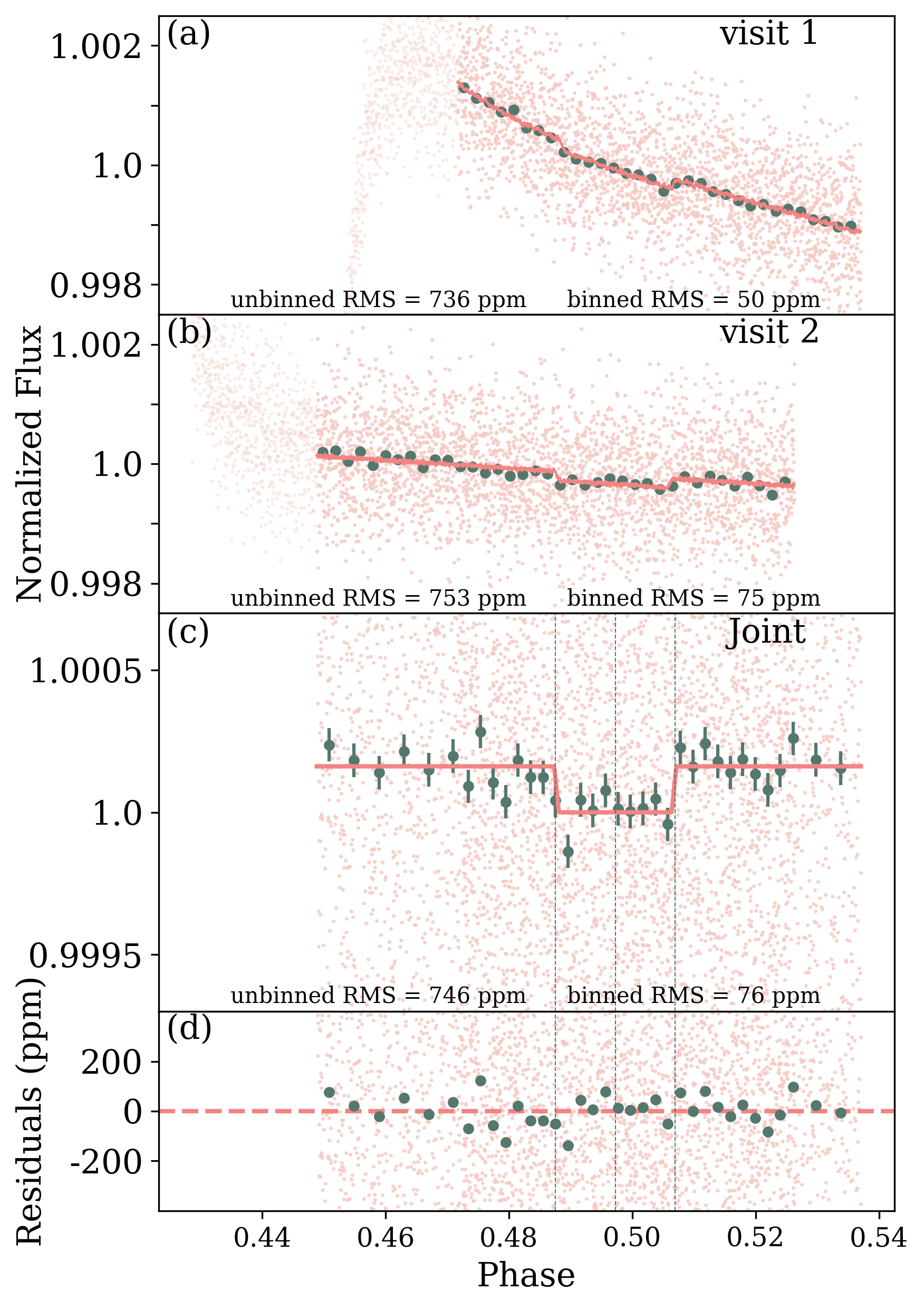}
    \caption{\textbf{(a) and (b)}: JWST light curves overplotted with the best-fit models using parameters listed in Table~\ref{tab:JWSTfittingresults}. Data excluded from the fit are in a paler shade. To make the eclipse more easily visible, we binned the individual integrations (pink points) with a bin width of 80 integrations (green points). \textbf{(c) and (d)}: Phase-folded light curves with systematics divided out, overplotted with the best-fit eclipse model. Panel (d) shows the residuals relative to the best-fit model. The vertical dashed lines indicate the observed start, midpoint, and end of the eclipse.}
    \label{fig:lightcurve}
\end{figure}

We reduced and analyzed the data with \texttt{SPARTA}, an independent, end-to-end JWST pipeline \citep{kempton_reflective_2023, zhang_gj_2024}. SPARTA has been used extensively on MIRI/LRS data, and we updated it for MIRI imaging data for this project. A detailed, step-by-step description of \texttt{SPARTA} as applied to MIRI imaging data is provided in Appendix\,\ref{append:sparta}. \texttt{SPARTA} routinely delivers smaller residuals (and, consequently, smaller errors on measured parameters) for MIRI/LRS timeseries data than other pipelines \citep[e.g.][]{powell_2024,xue_jwst_2024}. We find that it also consistently outperforms other pipelines for the TRAPPIST-1b and c MIRI F1500W secondary eclipse observations that we have tested it on (nine total visits, see Appendix\,\ref{append:sparta}). \texttt{SPARTA} with the newly implemented modules for MIRI Imaging will be publicly released on GitHub\footnote{https://github.com/ideasrule/sparta}. We present these results not as the definitive answer for the GJ\,3929b data, but as our unique contribution to the Rocky Worlds DDT program. We expect and encourage other groups to reduce the data with their own pipelines. 

Starting from the uncalibrated data, we identified four pixels ([66,64], [66,65], [67,64], and [67,65]), each located four pixels away from the PSF centroid, that exhibit anomalously low count levels. These pixels were subsequently flagged as bad by \texttt{mask\_0067.fits} and excluded from further analysis. Additionally, the first few groups on the brightest pixels show non-linear behavior, known as the reset-switch charge decay (RSCD), while the last group shows a ``pull-down'' effect. Therefore, we excluded the first five groups and the last group from the ramp-fitting process.

Among the methods we evaluated, simple aperture extraction yielded the lowest scatter in the light curve residuals compared to both optimal extraction and the z-cut approach (see Appendix \ref{append:sparta}). We optimized the aperture (i.e., target flux region) and annulus (i.e., background flux region) sizes by testing aperture radii from 4 to 10\,px (in 1-px increments), inner annulus radii from 8 to 26\,px, and outer annulus radii from 10 to 36\,px. For visit 1, the combination that minimized the scatter in the residuals was an aperture radius of 4\,px, with an inner annulus of 26\,px and an outer annulus of 30\,px. For visit 2, we again used a 4\,px aperture, pairing it with an inner annulus of 10\,px and an outer annulus of 30\,px.

For the separate analyses of each visit, we fixed the orbital period, a/R$_\star$, and inclination to the best-fit values from our joint RV and transit analysis. For eccentricity and $\omega$, since the solution is consistent with a circular orbit at the 2$\sigma$ level, we adopt $e=0$ for these fits (a 95\% ``detection'' of nonzero eccentricity typically requires $e / \sigma_e > 2.45$; \citealt{lucy_sweedney}). We allowed $T_{sec}$ to vary freely between the start and the end of each visit, and the secondary eclipse depth $F_p/F_s$ to vary freely between 0 and 1000\,ppm. The effect of allowing a negative eclipse depth is discussed below. For the joint analysis of both visits, we imposed Gaussian priors on a/R$_\star$ and inclination with the mean and the standard deviation obtained from the RV and transit analysis results in Table~\ref{tab:GJ3929_joint_fit},  while again treating $t_{sec}$ as a free parameter within the observation window of visit 1.

\renewcommand{\arraystretch}{1.0}
\begin{table*}[] 
\centering
\begin{tabular}{
  >{\arraybackslash}m{1.5cm}  
  >{\arraybackslash}m{4cm}  
  >{\centering\arraybackslash}m{5cm}  
  >{\centering\arraybackslash}m{5cm}  
}
\hline
& \textbf{Parameter} & \textbf{Prior} & \textbf{Posterior} \\
\hline

\multirow{6}{*}{\textbf{visit1}} 
& Period (d) & fixed & 2.6162644 \\
& $a/R_\star$&fixed & 17.04\\
&$inc$ (degrees)&fixed&89.30\\
&$e$&fixed&0\\
& $T_{sec} - 2400000.5$ (BJD)  & $\mathcal{U}(60864.387, 60864.558)$&60864.4541$^{+0.0016}_{-0.0021}$
\\
& $F_p/F_s$ (ppm) & $\mathcal{U}(0,1000)$& 177$^{+47}_{-45}$\\

\hline

\multirow{6}{*}{\textbf{visit2}}
& Period (d) & fixed & 2.6162644 \\
& $a/R_\star$&fixed & 17.04\\
&$inc$ (degrees)&fixed&89.30\\
&$e$&fixed&0\\
& $T_{sec} - 2400000.5$ (BJD)  & $\mathcal{U}(60887.874, 60888.077)$&60887.9993$^{+0.0014}_{-0.0050}$
\\
& $F_p/F_s$ (ppm) & $\mathcal{U}(0,1000)$& 143$^{+34}_{-35}$\\
\hline
\multirow{6}{*}{\textbf{joint fit}}
& Period (d) & fixed & 2.6162644 \\
& $a/R_\star$&$\mathcal{N}(17.04,0.5)$ & 16.8$^{+0.45}_{-0.42}$\\
&$inc$ (degrees)&$\mathcal{N}(89.3,0.5)$&89.50$^{+0.42}_{-0.38}$\\
&$e$&fixed&0\\
& $T_{sec} - 2400000.5$ (BJD)  & $\mathcal{U}(60864.387, 60864.558)$&60864.4537$^{+0.0009}_{-0.0014}$
\\
& $F_p/F_s$ (ppm) & $\mathcal{U}(0,1000)$& 160$^{+26}_{-27}$\\
\hline
\end{tabular} \caption{Priors and posteriors of the fit to the JWST secondary eclipse data. The model corresponding to the best-fit parameters is shown in Figure~\ref{fig:lightcurve}.}
\label{tab:JWSTfittingresults}
\end{table*}

We discarded the first 800 integrations ($\sim$76\,mins) of both visit 1 and 2 due to the ramps (which have very different shapes, see Figure~\ref{fig:lightcurve}) at the beginning of the observations. We modeled the light curve using the product of the eclipse model from \texttt{batman} and a systematics model:
\begin{equation}
\begin{split}
    F_{sys,i} &= F_{star}(1+m(t_i-t_{mean})+ \\
& \quad Ae^{-(t_i-t_0)/\tau}+c_x\Delta x_i+c_y\Delta y_i),
\end{split}
\end{equation}
where we define $\Delta x_i$ and $\Delta y_i$ as the residuals of the PSF centroid coordinates $x$ and $y$ at the $i$-th integration, calculated as the difference between the measured coordinates and their corresponding linear fits over time. Removing the $\Delta x_i$ and/or $\Delta y_i$ terms leaves F$_p/$F$_s$ unchanged, but slightly worsens the fit (higher Bayesian Information Criterion, BIC, by $<$6). Though adding linear terms of the PSF Gaussian width $c_{\sigma_x}\sigma_x$ and $c_{\sigma_y}\sigma_y$ results in similar BIC, the PSF widths are strongly correlated with the flux (e.g. $\sigma_x$ and $\sigma_y$ exhibit early-time ramps of the same shape as the flux) due to the ``brighter-fatter effect'' \citep{brighter_fatter_argyrious_2023}. We thus do not consider these two terms in the systematics model, to avoid removing astrophysical signals in the fit.

For the second visit, after excluding the first 800 integrations, we found that a linear trend is good enough to fit the data. The best-fit exponential time scale $\tau$ is 12 hr, which is twice the observation length. The exclusion of the exponential term from the systematics model for visit 2 led to a reduction in BIC by 13, indicating a statistical preference for the simpler linear model. For the first visit, however, the inclusion of an exponential ramp to the linear baseline is strongly favored: it reduces the BIC by 18 relative to a linear‐only model.

We summarize the median and $\pm1\sigma$ values of the free parameters inferred from our fit in Table~\ref{tab:JWSTfittingresults}. We find that the value of the error rescaling factor, a parameter in \texttt{SPARTA} to make the computed uncertainties consistent with the observed scatter in the residuals of the light curves, to be 1.16 for visit 1 and 1.18 for visit 2. We also report the extracted electron count rate (e$^-$/group) to be 8\% higher than the JWST Exposure Time Calculator (ETC)\footnote{https://jwst.etc.stsci.edu/} prediction for the same observation and aperture extraction strategy.

\textit{Is the eclipse actually there?} To answer this question,  we performed two additional tests beyond the results presented in Table~\ref{tab:JWSTfittingresults}. First, we allowed F$_p$/F$_s$ to vary between $-$1000 to 1000\,ppm (i.e., allowing an unphysical, negative eclipse). This setup enabled the sampling walkers to fully explore the parameter space and locate the minima without prior constraints. Keeping all other parameters and systematic model unchanged, for visit 1, we retrieved F$_p$/F$_s$ = $177^{+44}_{-46}$\,ppm (4$\sigma$ detection) with t$_{sec} = 2460864.9541^{+0.0016}_{-0.0021}$. For visit 2, we retrieved F$_p$/F$_s$= $138^{+37}_{-43}$\,ppm (3$\sigma$ detection) with t$_{sec} = 2460888.49942^{+0.0016}_{-0.0051}$. For the joint fit, we retrieved F$_p$/F$_s$ = $160\pm26$\,ppm (6$\sigma$ detection) with t$_{sec} = 2460864.9537^{+0.0010}_{-0.0014}$. In all cases, the F$_p$/F$_s$ values obtained when allowing negative depths are consistent within 1$\sigma$ with those reported in  Table~\ref{tab:JWSTfittingresults}.
Second, we compared the BIC of our best-fit eclipse model against a model with only systematics and no eclipse. We see that for visit 1, $\Delta {\mathrm{BIC}} =-32$. For visit 2, $\Delta \mathrm{BIC} = -21$. For the joint fit, $\Delta \mathrm{BIC} = -61$. We thus concluded that for all visits, eclipses are strongly favored over flat lines. 

By combining the mid-eclipse time (corrected for the light travel time effect) with the mid-transit time and orbital period, we constrain $e$\,cos\,$\omega$ using T$_\mathrm{sec}$ = T$_0 + \frac{P}{2}(1+\frac{4}{\pi} e\,\mathrm{cos}\,\omega)$. With the best-fit and $\pm1\sigma$ values of t$_{sec}$ from our fit to the JWST eclipse data, and the chain from the joint RV and transit analysis, we derived the $3\sigma$ upper limit on $e$\,cos\,$\omega$ of -0.0006 from visit1 and -0.002 from visit2. Thus we conclude that the JWST secondary eclipse times are indicative of a small, but non-zero eccentricity.

\subsection{Absolute stellar flux calibration}
Since the eclipse depth $F_p/F_s$ represents the planet-to-star flux ratio, accurate knowledge of the stellar flux across the F1500W bandpass is essential for determining the planet's dayside emission. In addition, stellar models play a key role in modeling the planetary surface and atmospheric emission spectra. To evaluate the reliability of the M-dwarf stellar models that we use in \S\ref{sec:model} \citep[\texttt{SPHINX} models,][]{iyer_sphinx_2023,iyer_2023_zenodo}, we performed absolute stellar flux measurement with \texttt{SPARTA} on GJ\,3929 using the JWST data (described in detail in Appendix~\ref{append:sparta}).

We first validated our code by analyzing all the available MIRI F1500W flux calibration data sets and comparing the results to the \texttt{CALSPEC} spectra\footnote{https://www.stsci.edu/hst/instrumentation/reference-data-for-calibration-and-tools/astronomical-catalogs/calspec} \citep{bohlin_techniques_2014, bohlin_new_2020,bohlin_how_2022, bohlin_faint_2024}, which is a database of HST and JWST standard stars. The comparison is summarized in Table~\ref{tab:append_stars}. \texttt{SPARTA} achieved better than 5\% agreement with \texttt{CALSPEC} for all 22 calibrators, demonstrating its robust performance.

With this validation complete, we proceeded to extract the absolute stellar flux of GJ\,3929 from the JWST observations. Taking into account the uncertainty in the stellar radius, distance to the star, and photon noise, we report an empirical absolute flux of 1,241$\pm 79$\,W\,m$^{-2}\,\mu$m$^{-1}$ from \texttt{SPARTA} at the surface of the star. Our measured flux is 4.8\% fainter than the SPHINX model for the adopted stellar parameters.  In comparison, \citet{ducrot_combined_2024} measured a TRAPPIST-1 flux 7\% lower than predicted by SPHINX, \citet{zieba_no_2023} reported fluxes within 1\% of the model, and \citet{fortune_hot_2025} found that SPHINX overpredicts the LHS\,1140 flux by 10\%.

\section{Atmosphere and surface modeling}\label{sec:model}

\subsection{Forward modeling}
\begin{figure*}[!t]
    \centering
    \includegraphics[width=1.0\linewidth]{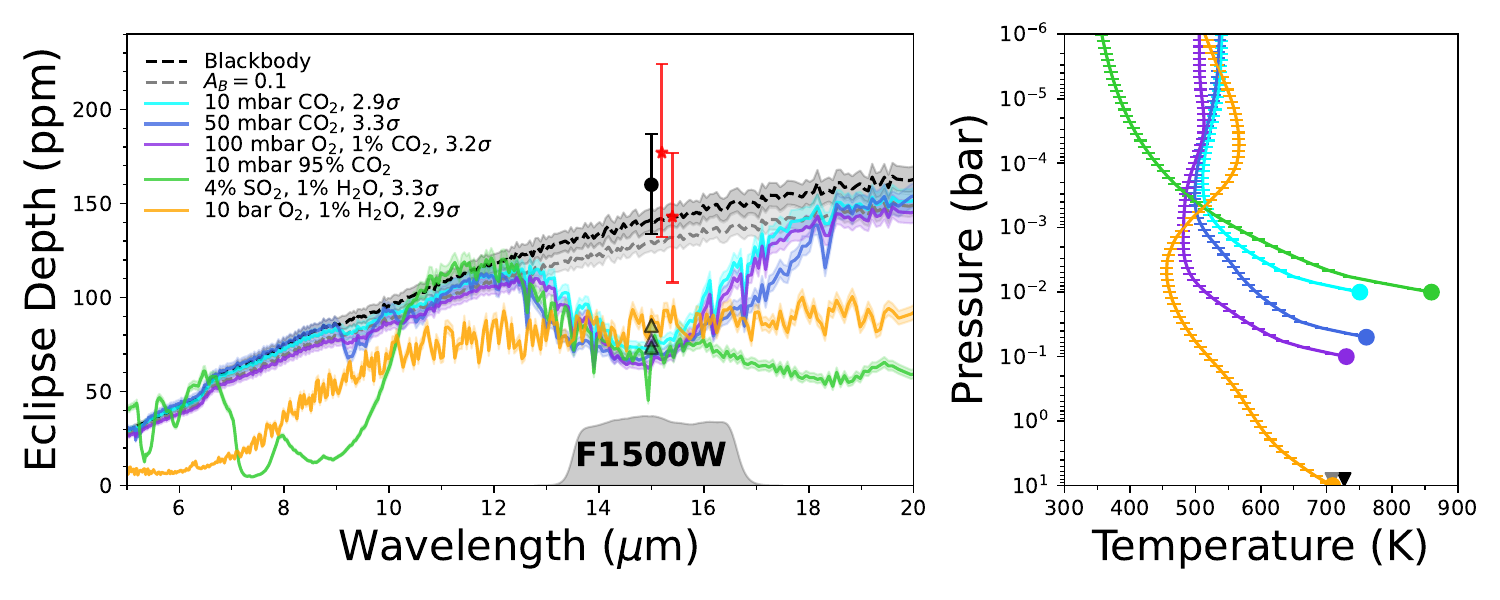}
    \caption{\textbf{(left)} Select \texttt{HELIOS} emission spectra for different atmospheric scenarios.  Shaded regions indicate 1$\,\sigma$ uncertainties due to the planet-to-star radius ratio $R_{p}/R_{\star}$. The black point represents our joint-fit eclipse depth, whereas the red errorbars are the individual visits (Table \ref{tab:GJ3929_joint_fit}).  Triangles represent the F1500W bandpass-integrated model eclipse depths. \textbf{(right)} \texttt{HELIOS}-derived temperature-pressure profile for each atmospheric scenario, with individual atmospheric layers indicated by horizontal lines. The disk-averaged surface temperatures for bare rock surfaces are indicated by triangles.  SO$_2$ acts as extreme greenhouse gases, efficiently warming the lower atmosphere to high temperatures. Our results rule out a variety of CO$_2$-rich atmospheres thicker than 100 mbar.}
    \label{fig:atmospheres}
\end{figure*}

To investigate whether GJ\,3929b hosts a thick atmosphere, we compare the measured eclipse depths with synthetic emission spectra computed using the radiative-convective equilibrium code \texttt{HELIOS 3.0} \citep{malik_helios_2017,malik_self-luminous_2019,malik_analyzing_2019,whittaker_detectability_2022}. \texttt{HELIOS} self‑consistently determines the 1D atmospheric temperature–pressure profile by balancing radiative and convective energy transport, and has been used in many similar investigations of rocky planet thermal emission spectra (e.g., \citealt{whittaker_detectability_2022,xue_jwst_2024,weiner_mansfield_no_2024,luque_dark_2025,coy2025,valdes_hot_2025,fortune_hot_2025}). We estimate the heat redistribution using the semi‑analytic scaling law of  \citet{koll_scaling_2022} for each scenario based on the short- and long-wave opacity of the atmosphere. We use a stellar spectrum interpolated from the SPHINX models \citep{iyer_sphinx_2023,iyer_2023_zenodo} with $T_{\mathrm{eff}}=3369$\,K, log\,\textit{g}=4.84, and [M/H]=0.00 \citep{kemmer_discovery_2022}. For planetary parameters, we adopted the median parameters presented in Table~\ref{tab:GJ3929_joint_fit}.\\ 

\textit{Atmosphere scenarios:}
We consider well-mixed atmospheres consisting of CO$_2$, H$_2$O, O$_2$, and/or SO$_2$. Molecular opacities were computed with HELIOS-K \citep{grimm2015helios}, with line lists for CO$_2$ from the HITEMP2010 database \citep{rothman2010hitemp}, H$_2$O from the ExoMol database \citep{polyansky2018exomol}, O$_2$ from HITRAN2020 \citep{gordon2022hitran2020}, and SO$_2$ from the ExoMol database \citep{underwood2016exomol}.  We employ k-distribution tables calculated at the resolution of the interpolated SPHINX spectrum ($R=250$ over 0.1--20$\,\mu$m and $R=50$ over 20--100$\,\mu$m) using 20 Gaussian points per sampling point.   
For all atmospheres, we assume a surface Bond albedo of 0.1.   CO$_2$ and H$_2$O in particular are expected to be ubiquitous in the secondary atmospheres formed by oxidized mantles like Earth and Venus \citep{LICHTENBERG2025}.  SO$_2$ is expected to outgas from highly-oxidized mantles, and has been proposed as a tracer for runaway internal melting due to tidal heating \citep{seligman_potential_2024}. While we do not consider the potential impacts of clouds, clouds are expected to increase Bond albedo and reduce the observed eclipse depth relative to a blackbody \citep{mansfield_identifying_2019}, inconsistent with our results.\\

\textit{No atmosphere cases:}
As the MIRI F1500W data alone are not sufficient to distinguish between surface types (e.g., \citealt{hu12,hammond_2025,paragas25}), we consider simple models of Bond albedo 0.0 and 0.1.  For reference, the Bond albedos of the Solar System airless bodies Mercury and the Moon are estimated at 0.088 and 0.136, respectively \citep{mallama2017spherical,matthews2008celestial}.\\

We show select atmosphere and surface models in Figure~\ref{fig:atmospheres}.  As the F1500W wavelength range covers a prominent CO$_2$ absorption feature, our data are particularly sensitive to even thin CO$_2$-rich atmospheres.  With just two eclipses, we are able to rule out a 50 mbar pure CO$_2$ atmosphere at 3.3$\,\sigma$.  A more pessimistic, still CO$_2$-rich composition of 100 mbar O$_2$ with 1\% CO$_2$ is also confidently ruled out at 3.2$\,\sigma$.  More exotic compositions, for example 10 bar of O$_2$ with 1$\%$ H$_2$O and 10 mbar of a SO$_2$-rich atmosphere (95\% CO$_2$, 4\% SO$_2$, and 1\% H$_2$O), are ruled out at 2.9$\,\sigma$ and 3.3$\,\sigma$, respectively. Although thin atmospheres (particularly those with H$_2$O as the only longwave absorber) are not ruled out entirely, our data are most consistent with a blackbody-like spectrum, expected of a bare rock.  While a deep eclipse depth may be explained by a strong thermal inversion in a CO$_2$-rich atmosphere, potentially induced by hydrocarbon hazes \citep{ducrot_combined_2024}, this scenario likely requires significant fine-tuning of haze parameters \citep{coy2025}, and we find that GJ\,3929b has likely experienced complete atmospheric erosion over its lifetime (\S\ref{sec:discussion}).

\subsection{Inferred Albedo and Heat Recirculation Efficiency}

Using our joint-fit eclipse depth, we fit for a dayside-integrated brightness temperature following methods outlined in \citet{coy2025}, taking into account uncertainties in $R_{p}/R_{\star}$, $a/R_{\star}$, [M/H], and $T_{\star,\mathrm{eff}}$ (\citealt{kemmer_discovery_2022} and our  Table~\ref{tab:GJ3929_joint_fit}). We find a dayside brightness temperature of 782$\pm79$\,K using SPHINX model stellar spectra, and 745$\pm76$\,K using the observed stellar flux, showing the insensitivity of our result to the choice of stellar model at the current measurement precision. For comparison, the dayside brightness temperature of a zero Bond albedo, zero heat distribution blackbody would be T$_{day,max}=737\pm 14$\, K. Following the definition of $\mathcal{R}$ (the `brightness temperature ratio' compared to a zero-albedo blackbody) in \citet{xue_jwst_2024}, we retrieved $\mathcal{R} = 1.07\pm0.10$ using SPHINX models, and $1.01\pm0.10$ using the observed flux. 

\begin{figure}[!t]
    \centering
    \includegraphics[width=1.0\linewidth]{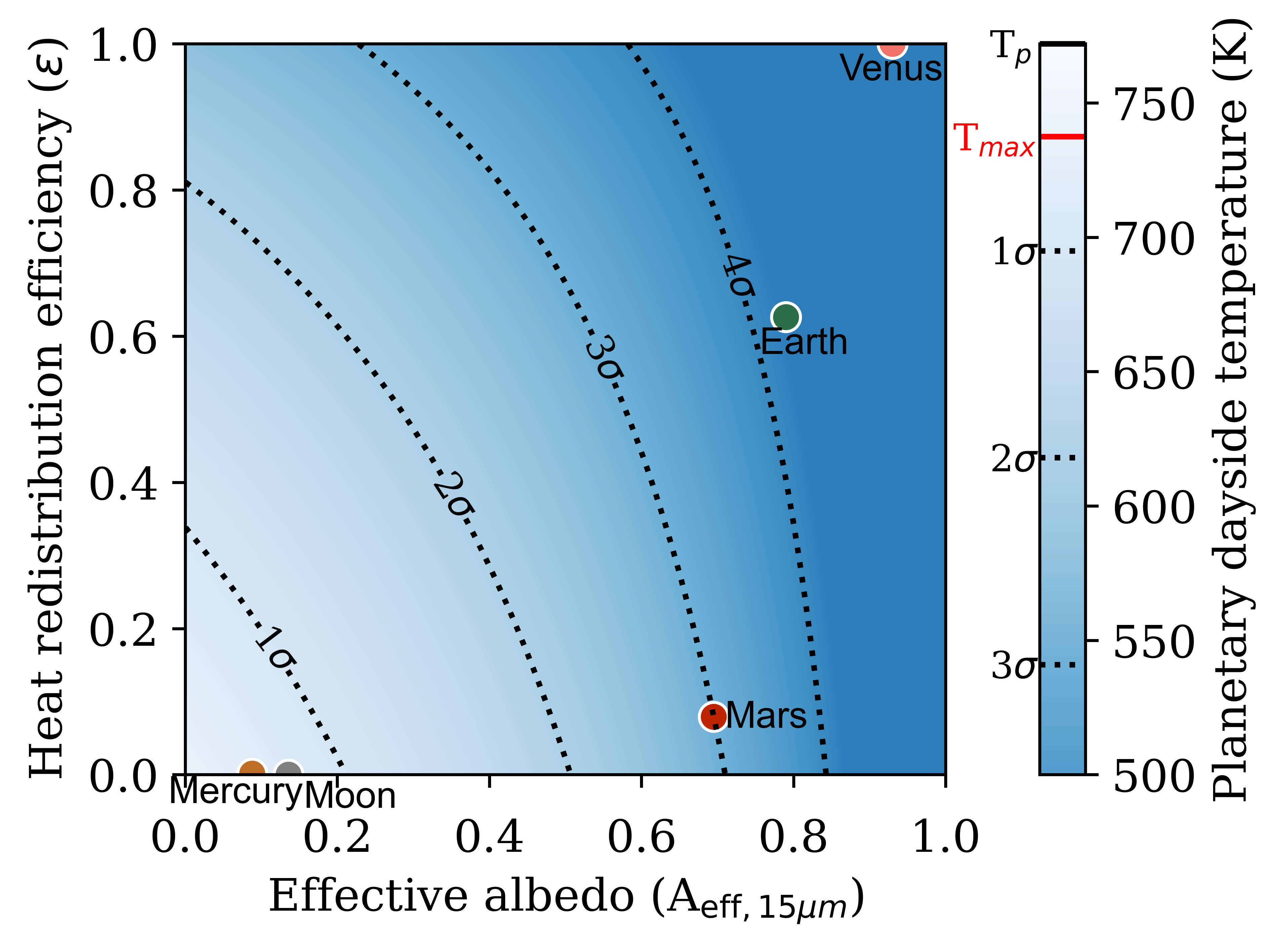}
    \caption{Planetary dayside temperature of GJ\,3929b as a function of effective albedo at 15 $\mu$m and heat redistribution efficiency compared to estimated values for Solar System bodies. For airless Mercury and Moon, we adopted the Bond albedo from \citet{mallama2017spherical} and \citet{matthews2008celestial} as their $A_{\mathrm{eff},15\mu m}$. For Mars, Earth, and Venus, since their atmospheres contain CO$_2$, we compute $A_{\mathrm{eff},15\mu m}$ from the modeled brightness-temperature spectra of Venus and Earth \citep[their Figure~1]{miller-ricci_emergent_2009} and the observed Mars spectrum \citep[their Figure~23]{tinetti_spectroscopy_2013}, averaging over the MIRI/F1500W bandpass using the filter throughput. $A_{\mathrm{eff}, 15\mu m}$ is then solved using $T_{p, 15\mu m} = \frac{T_{eff}}{ \sqrt{a/R_{\star}}}(1-A_{\mathrm{eff}, 15\mu m})^{1/4}(\frac{2}{3} - \frac{5}{12}\epsilon)^{1/4} $, where the heat-redistribution efficiency $\epsilon$ is estimated from the equator–pole temperature contrast following \citet[their Figure~4]{xue_jwst_2024}, since these planets are not tidally locked. We plotted the lower 1 to 4 $\sigma$ values of the inferred planet dayside temperature (T$_p$) of GJ\,3929b for better visualization. The median $T_p$ lies (slightly) above $T_{\mathrm{day,max}}$ and therefore falls outside the plotted range, but its value is indicated on the colorbar.}
    \label{fig:heatredist}
\end{figure}

In Figure~\ref{fig:heatredist}, we plot the planet's dayside temperature at 15\,$\mu$m as a function of effective albedo and heat redistribution efficiency, following equation 4 in \citet{cowan_statistics_2011}. With the measured planet's dayside temperature alone, Mars' atmosphere is disfavored at a 3$\sigma$ level, while Earth's and Venus' atmospheres are disfavored at $>3\sigma$ as such atmospheres would induce significant heat transport and lower the measured dayside emission (see Figure~\ref{fig:heatredist}). We caution readers that a planet with a Mars-like atmosphere orbiting an M dwarf receiving the same instellation as GJ\,3929b would certainly look different from Mars itself, due to a different T-P structure, and the points in Figure~\ref{fig:heatredist} represent the actual planets themselves, rather than generic ``Mars-like,'' ``Earth-like,'' or ``Venus-like'' atmospheres. However, from the standpoint of forward modeling, a Mars-like atmosphere (10\,mbar CO$_2$) is disfavored at the $2.9\sigma$ level, as shown in Figure~\ref{fig:atmospheres}.  For an airless body, we can infer a 2$\sigma$ upper limit on the effective albedo of $A_{\mathrm{eff}} = 0.49$, in agreement with a dark, bare rock.

\section{Discussion}\label{sec:discussion}

\subsection{Cosmic Shoreline}
\begin{figure} 
     \centering 
     \includegraphics[width=0.99\linewidth]{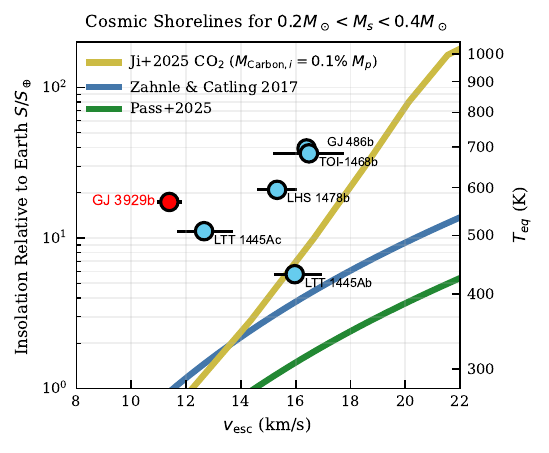}
     \caption{GJ\,3929b in context of the cosmic shoreline. The blue line represents the original XUV-driven shoreline proposed by \citet{zahnle_cosmic_2017}, with cumulative XUV flux converted to bolometric following their Eq. 27 with a stellar mass of 0.3 $M_\odot$. The green line is based on Eq. 10 of \citet{pass_receding_2025}, using a refined cumulative XUV history and PMS/flare corrections. The yellow line represents 90\% retention probability shoreline from \citet{ji_cosmic_2025}; estimated from hydrodynamic escape of CO$_2$ atmosphere, assuming an initial carbon inventory of 0.1 wt\% planetary mass. The planets in the comparison are the ones with JWST measurements in emission with host star mass from 0.2 to 0.4\,M$_\odot$}
     \label{fig:coshoreline_xuan}
 \end{figure}

Our forward modeling results support the conclusion that GJ\,3929b lacks a substantial atmosphere. This adds a valuable data point to evaluate the cosmic shoreline hypothesis, which posits that secondary atmosphere presence/absence is largely controlled by the cumulative XUV radiation received by a planet over its lifetime \citep{zahnle_cosmic_2017,pass_receding_2025, ji_cosmic_2025,berta-thompson_3d_2025}. We compare GJ\,3929b to three versions of the shoreline (Figure~\ref{fig:coshoreline_xuan}): the original XUV-driven loss shoreline proposed by \citet{zahnle_cosmic_2017}; the updated shoreline for fully-convective M dwarfs from \citet{pass_receding_2025}, which synthesizes observations to estimate the XUV history and incorporates flare corrections; and the \citet{ji_cosmic_2025} model tailored for CO$_2$-dominated atmospheres. The latter considers an initial carbon inventory of 0.1\% planetary mass, an optimistic estimate that is roughly ten times bulk silicate earth's carbon budget, and predicts a 90\% probability of atmospheric retention. Since CO$_2$ is the primary molecule probed at 15\,$\mu$m, this comparison directly informs the interpretation of our non-detection. GJ\,3929b lies well beyond all three shorelines we consider here, indicating it resides in a regime where atmospheric erosion is expected to outpace replenishment. Moreover, it is farther away from the shorelines than the bare rock planets TRAPPIST-1b and c, GJ\,1132b, and GJ\,486b ($<$0.2\,M$_\odot$ M-dwarf planets are not shown in the figure), making it the least likely to retain a CO$_2$-dominated atmosphere.

\subsection{Tidal heating}
Tidal dissipation is sometimes invoked to explain elevated surface temperatures on short-period planets, since an eccentric orbit can be periodically distorted by the host star (or companion planets), converting orbital energy into heat and, in extreme cases, sustaining magma oceans \citep{seligman_potential_2024}. However, our RV and transit analysis constrain GJ\,3929b’s eccentricity to $e=0.043^{+0.030}_{-0.021}$. Following \citep{coy2025}, we estimate that such an orbit could generate 280\,W\,m$^{-2}$ of tidal heating flux using a fixed-$Q$ model (assuming Earth-like tidal parameters, $Q=100$ and $k_{2}=0.3$) corresponding to an increase in $\mathcal{R}$ of only $\sim0.006$. This level of heating is insufficient to measurably raise the planet’s dayside brightness temperature, and we therefore rule out tidal heating as a significant contributor to the observed dayside brightness temperature.

\subsection{Thermal Beaming from Rough Surfaces}
While not statistically significant, our retrieved brightness temperature (782$\pm79\,$K) is slightly higher than that expected of a low-albedo blackbody (718$\pm$14$\,$K for $A_{B}=0.1$).  This has also been seen in two other planets with F1500W eclipse observations \citep{valdes_hot_2025,fortune_hot_2025}. One possible way to have $\mathcal{R}>$1 is \textit{thermal beaming}, a geometric effect that causes rough surfaces to appear hotter at opposition (e.g. \citep{davidsson2015interpretation,coy2025}.  This effect can lead to up to $\sim25\%$ increased thermal emission flux at 15$\mu$m during eclipse (B. P. Coy et al., in prep).  However, direct characterization of surface roughness likely requires high-precision full-orbit phase curves \citep{tenthoff24}, something likely out-of-reach for a long-period planet with a modest ESM like GJ\,3929b.

\section{Conclusion} \label{sec:conclusion}
The Rocky Worlds DDT program’s first two visits of GJ\,3929b yield no evidence for a thick CO$_2$- or H$_2$O–rich atmosphere, instead pointing to a dark, bare‐rock world. Leveraging our end-to-end  pipeline \texttt{SPARTA}, which performs independent data reduction, aperture photometry, absolute flux calibration, and rigorous systematics removal, we measure a secondary eclipse depth of 160$^{+26}_{-27}$\,ppm, and a dayside brightness temperature of 782$\pm79$\,K. This dayside temperature matches the zero–Bond-albedo, no-redistribution limit of T$_{day,max}\sim$737\,K. Atmospheric forward models show that even a 100\,mbar CO$_2$-dominated atmosphere is ruled out at $>3\,\sigma$.


GJ 3929b lies well on the airless side of many proposed versions of the cosmic shoreline. This indicates that the planet has likely lost any primordial or replenished atmospheres. In comparison with the other presumed bare rocks, TRAPPIST-1b and c, GJ\,1132b, and GJ\,486b, GJ\,3929b is the farthest from the shoreline, underscoring that it is highly unlikely to sustain a CO$_2$ atmosphere. 


The MIRI 15\,$\mu$m eclipses of GJ\,3929b deliver not only the first direct measurement of its dayside emission but also a powerful demonstration of our capability to probe atmospheres, or confirm their absence, on rocky planets orbiting M dwarfs. By showing that two single photometric points already disfavor a wide range of atmospheres, this result lays the groundwork for the full Rocky Worlds DDT program. We look forward to the new observations scheduled for GJ\,3929b and other Rocky World DDT targets. Together, these data will build a statistical framework for atmospheric retention, surface composition, and XUV-driven erosion on these small planets, transforming our understanding of how they form, evolve, and either keep or lose their atmospheres in the intense radiation environments \citep{coy2025, berta-thompson_3d_2025, ih2025}.


\begin{acknowledgments}
This work is based in part on observations made with the NASA/ESA/CSA James Webb Space Telescope. The data were obtained from the Mikulski Archive for Space Telescopes at the Space Telescope Science Institute, which is operated by the Association of Universities for Research in Astronomy, Inc., under NASA contract NAS 5-03127 for JWST. These observations are associated with program DD 9235. We thank the team led by co-PIs N.\ Espinoza and H.\ Diamond-Lowe for developing their observing program with a zero-exclusive-access period.

The MAROON-X group acknowledges funding from the David and Lucile Packard Foundation, the Heising-Simons Foundation, the Gordon and Betty Moore Foundation, the Gemini Observatory, the NSF (award number 2108465), and NASA (grant number 80NSSC22K0117). The Gemini observations are associated with programs GN-2021A-Q-120, GN-2022A-Q-120, and GN-2022A-Q-218 (PI: J.~Bean)

M.Z.\ thanks the Heising-Simons Foundation for support through the 51 Pegasi b fellowship. C.P.-G.\ is supported by the E. Margaret Burbidge Prize Postdoctoral Fellowship from the Brinson Foundation. M.R.\ acknowledges financial support from the Natural Sciences and Engineering Research Council of Canada through a Postdoctoral Fellowship. R.L.\ is supported by NASA through the NASA Hubble Fellowship grant HST-HF2-51559.001-A awarded by the Space Telescope Science Institute, which is operated by the Association of Universities for Research in Astronomy, Inc., for NASA, under contract NAS5-26555. R.L.\ acknowledges financial support from the Severo Ochoa grant CEX2021-001131-S funded by MCIN/AEI/10.13039/501100011033 and is funded by the European Union (ERC, THIRSTEE, 101164189). Views and opinions expressed are however those of the author(s) only and do not necessarily reflect those of the European Union or the European Research Council. Neither the European Union nor the granting authority can be held responsible for them.
\end{acknowledgments}

\facilities{JWST(MIRI), Gemini(MAROON-X), TESS}

\newpage
\appendix
\section{Overview of \texttt{SPARTA}} \label{append:sparta}
In this work, we present an updated version of \texttt{SPARTA}, an end-to-end data reduction pipeline that is completely independent of other existing pipelines. \texttt{SPARTA} was originally developed for reducing MIRI/LRS data, and was later extended to support NIRSpec/G395H and NIRSpec/PRISM observations (\citealt{kempton_reflective_2023, xue_jwst_2024, zhang_gj_2024, barat_metal-poor_2025}, Xue et al., in prep). In this appendix, we introduce new aspects in \texttt{SPARTA} for processing MIRI imaging time-series observations. Below, we outline the key stages and their respective functions:

1. \texttt{calibrate.py} starts from the \texttt{uncal.fits} and performs a series of corrections to produce calibrated slope images. The first step addresses emission correction: MIRI imaging data often exhibit coherent noise caused by electromagnetic interference (EMI), with a prominent 10 Hz component in imaging mode \footnote{\url{https://www.stsci.edu/contents/news/jwst/2024/pipeline-news-miri-emicorr-step-is-now-available}}. To remove this, \texttt{SPARTA} uses the same algorithm as the JWST official pipeline \texttt{jwst}\footnote{https://github.com/spacetelescope/jwst} with the \texttt{jwst\_miri\_emicorr.asdf} reference file from CRDS. Next, a row-by-row background subtraction is applied: a central region containing the source is masked out, and the median value along each row is computed from the remaining pixels and subtracted. If the observation utilizes the FULL array, the mean value of the reference pixels (defined as the four-pixel-wide strips along each edge of the detector) is subtracted from each group. Currently, no reference pixel correction is made for subarray data, even when the subarray includes the left column of
reference pixels \citep{morrison_jwst_2023}. The pipeline then applies linearity correction on a per-pixel basis using the polynomial coefficients in \texttt{jwst\_miri\_linearity.fits}, followed by dark current subtraction using \texttt{jwst\_miri\_dark.fits}. The resulting data are multiplied by the pixel-dependent gain to convert from DN to electrons. Finally, the slope of each pixel in each integration is computed using an optimally weighted scheme that accounts for both photon and read noise, as described in detail by \citet{kempton_reflective_2023}. The final products are saved in \texttt{\_rateints.fits} files. If absolute stellar flux is desired, a flat-field correction using \texttt{jwst\_miri\_flat.fits} is applied before the calibrated file is saved. 

2. \texttt{ap\_extract.py} performs aperture photometry on the \texttt{\_rateints.fits} and saves the resulting light curves. It begins by cropping each integration to a fixed subarray. For each integration, we first clean the image using sigma clipping. We then subtract the median value from the image and determine the target centroid in two steps: (1) identify the peak source within a specified region using \texttt{DAOStarFinder} \texttt{DAOStarFinder} \citep{larry_bradley_2024_13989456}, and (2) refine using iterative moment-based centroiding. A 2D Gaussian is optionally fit to the source to extract the center x, y position and the FWHM. Circular apertures and annuli are used for flux extraction and background estimation, respectively. The variance is computed by adding two terms in quadrature: (1) the per-pixel standard deviation of the background summing up across the aperture; (2) the per-pixel noise estimation from up-the-ramp fitting summing up across the aperture.  \texttt{SPARTA} automatically tests a range of aperture and annulus size combinations and generates a diagnostic plot (e.g., Figure~\ref{fig:mad_heatmap}) showing the mean absolute deviation (MAD) as a function of these parameters. The plot can help the user determine the optimal aperture size and annulus size to use. The background-subtracted flux is computed by subtracting the mean background (scaled to the aperture area) from the total aperture flux. The script also calculates the per-integration photometric uncertainty by propagating source photon noise, background photon noise, and read noise, the latter extracted from a reference file (\texttt{jwst\_miri\_readnoise.fits}). Finally, all fluxes, centroids, widths, background levels, and errors are saved for further use.

Likewise, \texttt{optimal\_apextract.py} performs optimal photometry on the \texttt{\_rateints.fits}. The code supports three weighting profiles for the optimal extraction: a per-integration 2D Gaussian fit to the median frame shifted to each centroid, a median-derived empirical profile from the cube, and a “z-cut” profile that captures the complex PSF structure by masking pixels below a fractional flux threshold. For the ``z-cut'' method, we followed the method described in \cite{august_hot_2025}.

3. \texttt{extract\_eclipse.py} first reads in time-series data (including fluxes, uncertainties, and centroid positions) from step 2 and normalizes the fluxes by their median. A 5$\sigma$ sigma-clipping routine is applied to remove outliers, and the masked values are excluded from the flux, time, uncertainty, and centroid arrays. Users can choose whether to bin the light curve prior to fitting; if binning is applied, the fluxes are renormalized and uncertainties are propagated accordingly. The centroid positions (x and y) are detrended by fitting linear trends as a function of time to remove pointing-induced systematics. The processed light curve is then modeled using a joint systematics and astrophysical model. Users can specify in a configuration file the prior types, prior ranges for the auxiliary parameters (e.g., orbital period, flux ratio), systematics models, and fitting method (e.g., MCMC or nested sampling).

4. \texttt{get\_stellar\_flux.py} computes the absolute stellar flux from the \texttt{\_rateints.fits}. When calibrating absolute stellar flux, \texttt{emicorr} and  \texttt{gain\_scale} need to be skipped, and \texttt{flat\_field} needs to be turned on \citep{gordon_james_2025}. The script first performs aperture photometry using the aperture radius and annulus radius specified in the \texttt{jwst\_miri\_apcorr.fits}. After subtracting the sky background and summing the DN counts within the aperture, the DN counts are scaled by a conversion factor that accounts for the finite aperture size. This factor adjusts the measured flux to the expectation from an ``infinite" aperture.  Then the step applies instrument-specific photometric calibration using the scaling factor in \texttt{jwst\_miri\_photom.fits} to convert DN data into physical flux units (W/m$^2$/$\mu$m).

\section{Validating with TRAPPIST-1b \& c}
To verify the performance of our pipeline, we re-processed the published JWST/MIRI observations of TRAPPIST-1 b and c with \texttt{SPARTA} and compared the recovered eclipse depths with the literature values \citep{greene_thermal_2023,ducrot_combined_2024,zieba_no_2023}. For TRAPPIST-1\,b, we imposed a uniform prior on the eclipse mid-time, $t_{\rm eclipse}$, spanning $\pm0.1\,$days around the best‐fit value from \citet{greene_thermal_2023}. For TRAPPIST-1c, we did not include the group-level row-by-row background subtraction since it increased the scatter in the light curve by $\sim$10\%. We applied a Gaussian prior on $t_{\rm eclipse}$ with mean $\mu$ and standard deviation $\sigma$ set to the median and three times the uncertainty predicted by TTV analysis \citep{agol_refining_2021}. This choice was necessary because a uniform prior failed to yield a meaningful constraint on $t_{\rm eclipse}$ in visits 1 and 3. In Table~\ref{tab:sparta_validation}, we compare the eclipse depths recovered for TRAPPIST-1b and c by \texttt{SPARTA} and those reported in the literature. 

\begin{figure}[h!]
    \centering
    \includegraphics[width=0.9\linewidth]{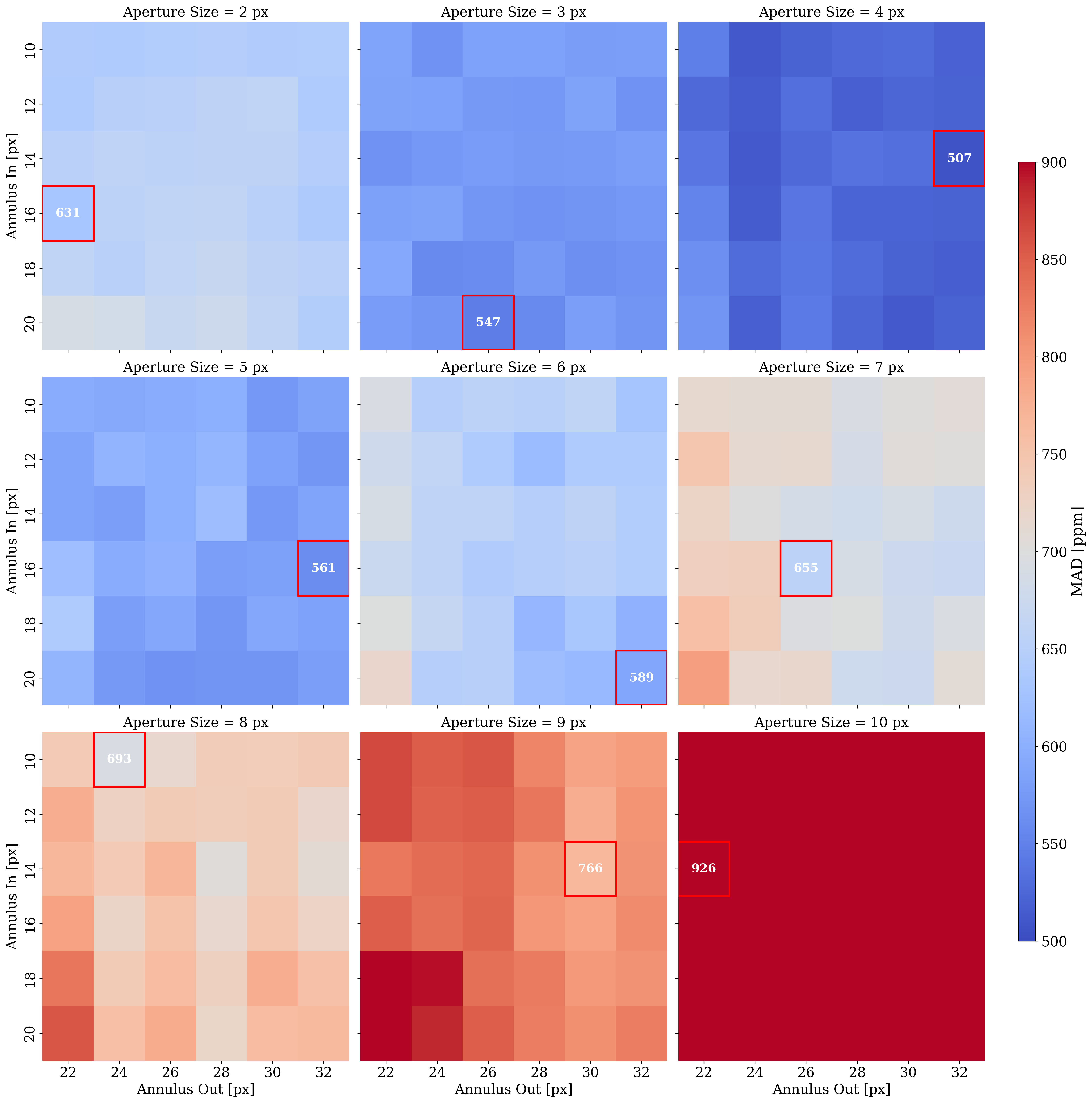}
    \caption{An example of \texttt{ap\_extract.py}'s diagnostic plot, showing the MAD of light curves as a function of aperture and annulus sizes.}
    \label{fig:mad_heatmap}
\end{figure}

When validating SPARTA on these datasets, we consistently saw smaller light curve scatter and tighter constraints on the secondary eclipse depths than all published analyses. We explored aperture radii from 2 to 10 pixels (in 1-pixel steps), inner-annulus widths from 10 to 22 pixels, and outer-annulus widths from 22 to 32 pixels (in 2-pixel steps). Across all configurations, a 4-pixel aperture radius yielded the minimum MAD in the light curve (see the heatmap for visit 1 of TRAPPIST-1 b in Figure~\ref{fig:mad_heatmap}).

\renewcommand{\arraystretch}{1.0}
\begin{table}[h!] \label{tab:sparta_validation}
\caption{Eclipse depths (in units of ppm) derived using \texttt{SPARTA} are compared against both literature values and results from the recently released \texttt{Erebus} pipeline, which employs FN-PCA \citep{connors_uniform_2025}. For TRAPPIST-1b, we adopt literature values from Extended Data Table 4 in \citet{ducrot_combined_2024}, selecting the measurements with the smallest reported uncertainties among the four independent data reductions and analyses. For TRAPPIST-1c, we adopt literature values from  the ED reduction in \citet{zieba_no_2023}}
\begin{tabular}{M{3cm} M{3cm} M{3cm} M{3cm} M{3cm}}

\hline
\multicolumn{1}{c}{Planet}  & visit \# & \texttt{SPARTA} & Literature & \texttt{Erebus} \\ \hline
\multirow{6}{*}{TRAPPIST-1b} 
& 1 &  799$\pm$154 &    771$\pm$176 &  780$\pm$191\\
& 2 & 680$\pm$138 & 691$\pm$169           & 1044$\pm$192 \\
& 3 & 574$\pm$148     & 950$\pm$170    & 599$\pm$217\\
& 4 & 751$\pm$170   & 759$\pm$176& 728$\pm$216 \\
& 5 & 803$\pm$152 &    790$\pm$164& 795$\pm$170\\
& joint  & 748$\pm$65 & 740$\pm$80 & 863$\pm$90       \\ \hline
\multirow{5}{*}{TRAPPIST-1c} 
& 1        &314$\pm$144 & 445$\pm$193 &-131$\pm$211 \\
& 2        & 474$\pm$154& 418$\pm$173 &301$\pm$202 \\
& 3        &367$\pm$159 & 474$\pm$158 &335$\pm$199        \\
& 4        & 402$\pm$166 & 459$\pm$185 &1064$\pm$177        \\
& joint    & 329$\pm$79 & 423$^{+97}_{-95}$ &312$\pm$128 \\ \hline

\end{tabular}

\end{table}

\section{Validating with JWST calibrators}
We validated the absolute stellar flux extraction workflow in \texttt {SPARTA} with the JWST flux calibrators for all available MIRI F1500W data sets. The results are summarized in Table~\ref{tab:append_stars}. We compared \texttt{SPARTA}'s extracted flux to the reference \texttt{CALSPEC} spectra \citep{bohlin_techniques_2014,bohlin_new_2020,bohlin_how_2022,bohlin_faint_2024}. \texttt{CALSPEC} contains the composite stellar spectra that are flux standards on the HST and JWST systems. The \texttt{CALSPEC} spectra were weighted by the F1500W filter throughput computed using \texttt{pandeia}. We then calculated the fractional difference between the \texttt{SPARTA}-extracted fluxes and the weighted \texttt{CALSPEC} values. For context, we also report the corresponding differences between \texttt{CALSPEC} and the JWST calibration results from \citet{gordon_james_2025}. In nearly all cases, \texttt{SPARTA}'s fluxes agree with \texttt{CALSPEC} to within 5\%. The only exception is HD\,101452, which shows a discrepancy of approximately 6\%. However, this star also exhibits a similar offset in the JWST calibration paper, suggesting the deviation is likely due to the source itself rather than the \texttt{SPARTA} pipeline.

\begin{table}[h] \label{tab:append_stars}
\centering
\begin{tabular}{clcccc}
\hline
\multicolumn{1}{l}{}           & Star                    & Program & Subarray  & \texttt{SPARTA} (\%) & Ref (\%) \\ \hline
\multirow{9}{*}{Solar analogs} 
& 16 CygB                 & 1538    & SUB64     & -5.0     & -3.5      \\
& GSPC\,P177-D             & 1538    & BRIGHTSKY & 2.4    & 1.3       \\
& GSPC\,P330-E             & 1538    & BRIGHTSKY & 1.4    & -0.4      \\
&                         & 4498    & FULL      & -1.4   & 0.9       \\
& HD\,106252                & 1538    & SUB64     & -3.3   & 0.2       \\
& HD\,142331                & 4498    & SUB64     & 0.3    & 1.0         \\
& HD\,167060                & 1538    & SUB64     & -1.3   & -2.4      \\
& HD\,37962                 & 1538    & SUB64     & -2.8   & -0.3      \\
& HR\,6538                  & 4498    & SUB64     & -3.3   & 0.9       \\ \hline
\multirow{4}{*}{Hot Stars}  
& G\,191-B2B         &1537    &FULL & 3.0 &3.0 \\
& 10\,Lac                   & 4497    & SUB64     & 0.7    & 4.6       \\
& $\mu$ Col                  & 4497    & SUB64     & -0.3   & 4.0         \\
& $\lambda$ Lep              & 4497    & SUB64     & -0.9   & 2.8       \\ \hline
\multirow{14}{*}{A Dwarfs}     
& HD\,2811                  & 1536    & SUB64     & -2.9   & -1.1      \\
&                         & 4496    & BRIGHTSKY & -2.9   & -2.3      \\
& BD+60-1753              & 1027    & SUB256    & -3.3   & -2.4      \\
&                         &         & FULL      & -2.6   & -2.8      \\
&                         & 1536    & SUB256    & -1.9   & -2.0        \\
& del\,UMI                 & 1536    & SUB64     & -5.0     & -2.7      \\
& HD\,163466                & 1027    & SUB64     & -3.3   & -1.7      \\
&                         & 1536    & SUB64     & -3.6   & -1.6      \\
& HD\,5467                  & 4496    & SUB256    & -2.2   & -2.9      \\
& HD\,101452                & 4496    & BRIGHTSKY & -6.8   & -6.6      \\
& 2MASS\,J17430448+6655015 & 1027    & FULL      & -2.7   & -3.4      \\ 
& 2MASS\,J17571324+6703409 & 4496    & BRIGHTSKY & -2.1 & -3.3 \\
& 2MASS\,J18022716+6043356 & 1536    &BRIGHTSKY & -0.3 &-3.5 \\ 
& HD\,55677 & 4496 & SUB256 & -5.3 & -2.4 \\ \hline
\end{tabular} \caption{Comparison of absolute stellar fluxes extracted with \texttt{SPARTA} to reference \texttt{CALSPEC} spectra for a sample of JWST flux calibrators. The sample includes all calibrators observed with the F1500W filter to date, excluding HD\,180609. Although HD\,180609 is the only calibrator observed using both the SUB128 subarray and the F1500W filter—the same configuration as the GJ\,3929b observation presented in this work—it was later identified as a debris disk host \citep{gordon_james_2025} and is therefore unsuitable as a flux calibrator. The fourth column reports the fractional differences between the \texttt{SPARTA} fluxes and the \texttt{CALSPEC} predictions, calculated as $\frac{Flux_{\mathrm{SPARTA}} - Flux_{\mathrm{CALSPEC}}}{Flux_{\mathrm{CALSPEC}}}$. The fifth column reports the fractional differences between the fluxes reported in \citet{gordon_james_2025} and the \texttt{CALSPEC} predictions. \texttt{SPARTA} achieves better than 5\% agreement for all stars except HD\,101452, which also shows a significant discrepancy in the published JWST calibration.}
\end{table}

\bibliography{reference}{}
\bibliographystyle{aasjournalv7}



\end{document}